\documentclass[letter]{aa} % for the letters 
\usepackage{graphicx}
\usepackage{txfonts}
\usepackage{natbib}
\usepackage[caption=false]{subfig}
\usepackage[linktocpage]{hyperref}
\usepackage{xcolor}
\usepackage{threeparttable}
\usepackage{etoolbox}
\makeatletter
\patchcmd\@combinedblfloats{\box\@outputbox}{\unvbox\@outputbox}{}{%
\errmessage{\noexpand\@combinedblfloats could not be patched}%
}%
 \makeatother

\begin{document}

   \title{The mass-size plane of EAGLE galaxies}

    \author{M. S. Rosito \inst{1}, P. B. Tissera\inst{2,3,4}, S. E. Pedrosa\inst{1,5}, C. D. P. Lagos\inst{6,7}} 

   \institute{Instituto de Astronom\'{\i}a y F\'{\i}sica del Espacio, 
CONICET-UBA, Casilla de Correos 67, Suc. 28, 1428, Buenos Aires, Argentina.
         \and
             Departamento de Ciencias F\'{\i}sicas, 700 Fernandez Concha,  Universidad Andr\'{e}s 
Bello, Santiago, Chile. 
\and Corresponding Investigator, IATE-CONICET, Laprida 927, C\'ordoba, Argentina.
\and Millenium Institute of Astronomy, 700 Fernandez Concha,  Universidad Andr\'{e}s Bello, Santiago, Chile.
\and Departamento de F\'{\i}sica Te\'{o}rica, Facultad de Ciencias, Universidad Aut\'{o}noma de Madrid.
\and International Centre for Radio Astronomy Research (ICRAR), M468, University of Western Australia, 35 Stirling Hwy, Crawley, WA 6009, Australia.
\and ARC Centre of Excellence for All Sky Astrophysics in 3 Dimensions (ASTRO 3D).}
	\abstract
   {Current observational results show that both late-and-early-type galaxies follow tight mass-size planes, on which
physical properties such as age, velocity dispersion and metallicities correlate with the scatter on the plane. }
   {We study  the mass-size plane of  galaxies in cosmological hydrodynamical simulations,  as a function of velocity dispersion, age, chemical abundances, ellipticity and spin parameters with the aim
at assessing to what extent the current cosmological paradigm can reproduce these observations and provide a physical interpretation of them. }
   {We select a sample of well-resolved galaxies from the $(100~{\rm Mpc})^3$ simulation of the EAGLE Project. This sample is composed by 508 spheroid-dominated galaxies and 1213 disc-dominated galaxies.
The distributions of velocity dispersion, age, metallicity indicators and gradients and spin parameters across the mass-size plane are analysed. Furthermore, we study the relation between shape and kinematic parameters. The results are compared with observations.}
   {The mass-weighted  ages of the EAGLE galaxies are found to vary along lines of  constant velocity dispersion on the mass-size plane, except for
galaxies with velocity dispersion larger than $\sim 150 $km s$^{-1}$.
     Negative age gradients tend to be found in extended disc galaxies in  agreement with observations. However, the age distributions of  early-type galaxies show a larger fraction with 
     inverted radial profiles. The distribution of metallicity gradients does not show any clear  dependence on this plane.
     Galaxies with similar spin parameters ($\lambda$) display larger sizes as their dynamical masses increase.  Stellar-weighted ages are found to be good proxies for $\lambda$ in galaxies with
     low ellipticity ($\varepsilon$). A bimodal distribution of $\lambda$ is found so that the high-$\lambda$ peak is dominated by discs with young SPs while the second peak is mainly populated by
slow rotators ($\lambda <~0.2$) with old stars.
Our findings suggest that the physical processes which
regulate the star formation histories in galaxies might also affect the angular moment budgets of gas and stars
and as a consequence, their morphology.
 }
  {}


\authorrunning{M. S. Rosito et al.}
  \maketitle
%------------

\section{Introduction}

The formation and evolution of galaxies is a complex interplay of different physical processes such as rapid gas collapse, disc mergers, minor mergers and secular
evolution, among others. These processes contribute to regulate the star formation activity and also to mix the chemical elements, polluting the interstellar medium (ISM) of galaxies. The formation of spiral galaxies can be understood within the scenario of global conservation of specific angular momentum \citep{fall2013}. However, in a hierarchical universe, mergers and interactions as well as other environmental effects may redistribute the angular momentum of gas and stars, modifying the morphologies and overall properties of galaxies \citep[e.g.][]{Pedrosa2015,lagos2017, Lagos2018b}. Ellipticals are
expected to have a complex formation history where the mentioned processes might also take place, albeit acting with different efficiencies and characteristics \citep{clauwens2018}. In particular, the mass-size plane, which combines  two fundamental properties of galaxies, could hold relevant information for the understanding of galaxy formation \citep[][hereafter Li2018]{Li2018}.

The galaxy assembly process is expected to leave imprints on chemical patterns on the stellar populations (SPs). 
As baryons are transformed into stars, stellar evolution takes place and new chemical elements  are synthesised and ejected 
into the ISM, at different stages of evolution of the stellar progenitors. The new-born SPs affect the chemical 
abundances of the ISM, building up 
the chemical history of galaxies. The growth of stellar mass and galaxy size via star formation, gas accretion or/and 
galaxy mergers are therefore expected to be related to the properties of the SPs such as chemical abundances,  
kinematics and ages.

 With the recent help of integral-field spectroscopy (IFS) techniques, 
 galaxy surveys such as MaNGA \citep{Bundy2015},  CALIFA \citep{CALIFA2012} and SAMI \citep{bryant2015} 
 have gathered information which allows the construction of the mass-size plane populated 
 with enough galaxies to study the interdependencies on those properties with higher statistical 
 signals than it was possible before \citep[e.g.][]{Ryden2001}. Li2018 analysed the mass-size plane of 
 early-type galaxies (ETGs) and spiral galaxies (LTGs) in the  MaNGA survey, finding that both types of 
 galaxies are on tight mass-size planes, which are consistent with the predictions of 
 the virial theorem. Metallicities and ages are found to vary systematically on the mass-size plane, 
 along the direction of constant velocity dispersion. Galaxies of different morphologies  
 follow  mass-size planes with distinct dependences on other galaxy properties. Overall, Li2018 found that 
 the velocity dispersion may be used as a proxy for the bulge-to-total ($B/T$) ratio \citep[e.g.][]{CappellariAtlasXX}. In their sample, 
 ETGs of higher velocity dispersion are typically associated with older and higher metallicities SPs. 
 Additionally, \citet{sande2018} analysed the link between the rotation-to-dispersion velocity ratio of a galaxy 
 and its ellipticity, finding that stellar age follows ellipticity ($\varepsilon$) very well in oblate rotating spheroids.  
 \citet{graham2018} reported a bimodal distribution in the $(\lambda, \varepsilon)$ plane of galaxies in the MaNGA survey, with the secondary peak being
dominated by slow rotator galaxies.

Cosmological simulations that include chemical evolution are fundamental tools to study the relation between dynamical, 
structural and chemical properties \citep[e.g.][]{mosco2001, lia2002, tissera2012,taylor2017}.
These simulations show that the formation of LTGs could be   explained if  overall global angular momentum
conservation takes place. Hence, they mostly grow inside-out, leading to negative age and metallicity gradients \citep{Tissera2016, tissera2019}. These trends may be disturbed if processes such as  galaxy mergers, 
bars and migration take place. The formation of  ETGs could involve the action of a variety of mechanisms at different stages of their evolution \citep{Naab2013}, such as minor and major mergers, as well as early filamentary, cold gas accretion.
Indeed, galaxy mergers are considered the most efficient mechanisms to redistribute angular momentum and modify the morphologies of galaxies, among other galaxy properties \citep[e.g.][]{bois2011,perez2013,naab2014,Lagos2018b}.
\citet{lagos2018} found that ~70 per cent of the slow rotators have experienced at least one important merger event in the EAGLE simulations. They also found that wet mergers tend to increase the $\lambda$ while dry mergers have the opposite effect. Recently, \citet{sande2019} analysed the structural, dynamical and stellar age populations of galaxies
in several large-volume simulations, including the EAGLE simulations, and confront them with IFS observations. These authors reported a good match of the mass-size plane between observations and simulated galaxies but did not find a correlation between $\varepsilon$  and age that resembled  observations.

In this paper, we analyse the mass-size relation of simulated galaxies extracted from the EAGLE simulations.
The EAGLE simulations is able to reproduce relatively well the observed diversity of galaxy morphologies \citep[][and references therein]{Trayford2018}.
 We classify the analysed galaxies in spheroidal-dominated (hereafter, E-SDGs) and disc-dominated (E-DDGs) systems according to the fraction of their stellar mass that is rotationally supported.
In a previous work, \citet{Rosito2018b} studied the structural relations and the mass growth history (MGHs) of dispersion-dominated galaxies selected from the EAGLE project, finding
that they are able to reproduce structural relations such as  the mass-size relation and the fundamental plane.  The MGHs analysis of \citet{Rosito2018b} shows a trend for a coeval formation of the SPs in massive E-SDGs and for a weak outside-in formation  in low-mass E-SDGs, driven by a rejuvenating star formation activity that took place in the central regions.
On the other hand, the E-DDG  galaxies tend to form inside-out as expected.
We use 3D (real) and 2D (projected) distributions in order to analyse the intrinsic distributions as well as to compare with observations.

This paper is organised as follows. Section \ref{sec:simu} describes the main characteristics of the simulations. Section \ref{sec:ms} explores the mass-size plane as a function of galaxy properties.
Finally, Section\ref{sec:con} summarises our main findings.

\section{Simulated galaxies}
\label{sec:simu}

The EAGLE project is a suite of cosmological simulations\footnote{We use the publicly available database by \citet{mcalpine2016}.} run with a modified
version of {\small GADGET-3} code that includes radiative cooling, star formation, chemical evolution
and stellar and AGN feedback \citep{schaye2015}.
Details on the subgrid physics adopted to simulate these processes can
be found in \citet[][]{schaye2015},  \citet{Crain2015} and \citet{Rosas2016}.
The simulations are produced assuming a $\Lambda$ Cold Dark Matter scenario with the Planck Cosmology \citep{Planck2014a, Planck2014b}.
For this work, we use the 100 Mpc box-sized reference simulation that
allows us to explore a large statistical sample of galaxies with
different morphologies.

The classification of EAGLE  galaxies into E-DDGs and
E-SDG has been done based on a
dynamical criteria to define the rotational-and-dispersion-dominated stellar components (i.e. disc and bulge, respectively) as explained in \citet{tissera2019}. 
Hence, all galaxies have been rotated so that the plane of rotation is perpendicular to the total angular momentum vector. Galaxies are classified according to their $B/T$ ratio, which
directly quantifies the stellar mass fraction that is dominated by velocity dispersion. For consistency with \citet{Rosito2018b}, we adopt $B/T=0.5$ to separate disc (E-DDG) from
bulge (E-SDG) dominated galaxies.
The analysed sample comprises 508 E-SDGs and 1213 E-DDGs
resolved with more than 10,000 star particles within the optical
radius\footnote{The optical radius is defined as the one that encloses
~80 per cent of the baryonic mass of the galaxy.}. They are all central galaxies, which 
corresponds to the main galaxy in the potential well of a halo. Satellite galaxies are not analysed in this work.

We define the 3D half-mass radius $R_{\mathrm{hm}}$  as  the one that encloses 50 per cent of the
stellar mass of the simulated galaxies. 
For each simulated galaxy, the median value of chemical abundances  (O/H and [O/Fe]) are calculated within $R_{\mathrm{hm}}$. 
We also estimate the dynamical mass ($M_{\rm dyn}$), assuming virialization for both the
observations and the simulations, as shown in \citet{Rosito2018b}.

The chemical abundance and age profiles are estimated within three
times the gravitational softening (0.7 kpc) and $R_{\mathrm{hm}}$ and normalised by  $R_{\mathrm{hm}}$. Linear  regression fits (in logarithm space) are performed to the normalized profiles, following the procedure given by Li2018 (within the same
radial interval mentioned above).

We also estimated the projected half-mass radius, $R_{\mathrm{hm}}^{2\mathrm{D}}$.  The $\lambda$ and $\varepsilon$ parameters are taken from \citet{lagos2018}. 
These authors measured the r-band luminosity-weighted line-of-sight velocity  stellar $\lambda$
and $\varepsilon$ at the projected half-r-band luminosity radius, using random inclinations to best mimic observational procedures. 

\section{Analysis}
\label{sec:ms}

We explore the distribution of velocity dispersion, age, metallicity and $\lambda$ of the EAGLE galaxies on  the mass-size plane. 
It is important to bear in mind that observed and simulated samples have been constructed using different selection criteria. Li2018 use the S\'ersic index to distinguish ETGs and LTGs so that galaxies with $n_{\mathrm{Sersic}}>2.5$ are classified as early types. They also use the dynamical mass and the major semi-axis of the half-light isophote. For the selected  EAGLE sample, we use the dynamical $B/T$ ratio to distinguish morphologies and  the half-mass radius as the characteristic size, consistently with \citet{Rosito2018b}. We also use the mass-size relations for passive and active star-forming (SF) galaxies reported by \citet{wel2014}.  Passive SF galaxies are compared to E-SDGs while active SF are confronted with E-DDGs.

\begin{figure*}
  \centering
\includegraphics[width=0.9\textwidth]{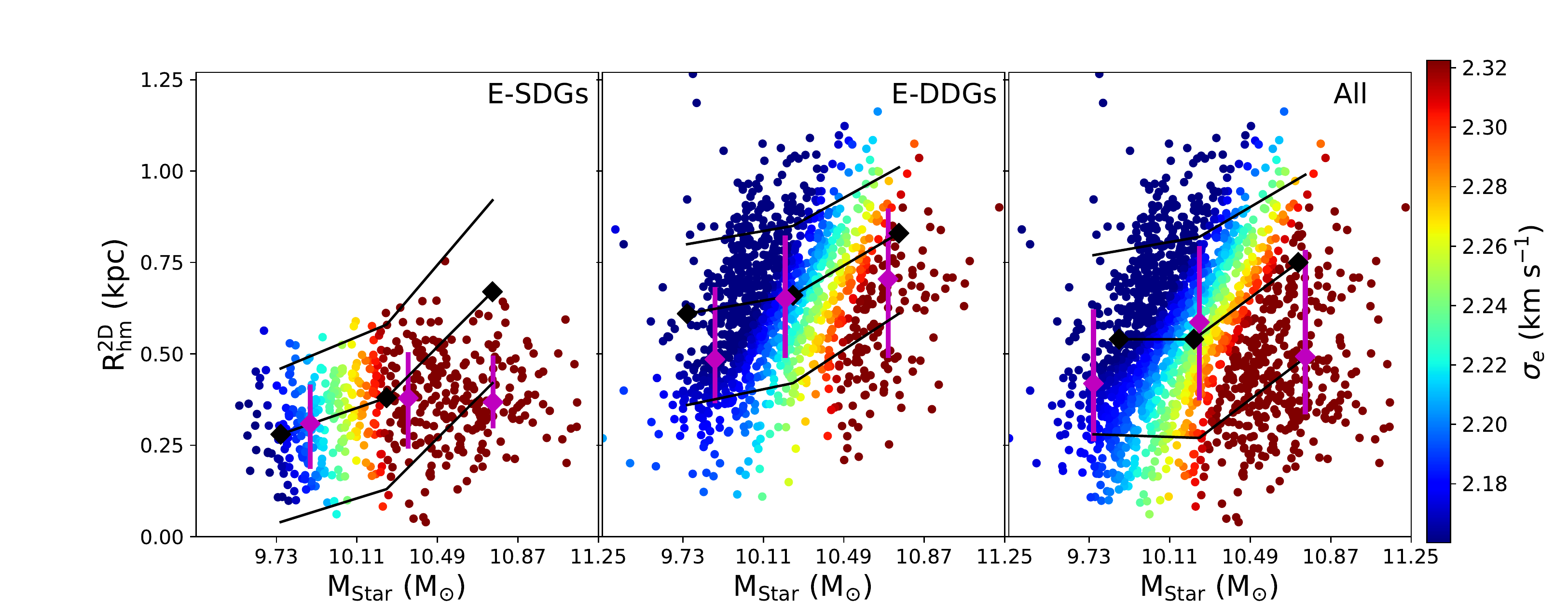}
  \caption{Velocity dispersion ($\sigma_e$) LOESS-smoothed distributions  on the stellar mass-size plane for E-SDGs (left panel), E-DDGs (middle
    panel) and all galaxies (right panel) in the EAGLE simulation at $z=0$. The median relations for the EAGLE galaxies (pink rhombus) are also shown.
For comparison the median relations for passive (left panel), active (middle panel) and all galaxies together (right panel) reported by \citet{wel2014} are also included in black rhombus. 
 Error bars correspond to the 16 and 84 percentiles  for both 
simulated and observed data.}
  \label{fig:fig1}
\end{figure*}

\begin{figure*}
  \centering
\includegraphics[width=0.9\textwidth]{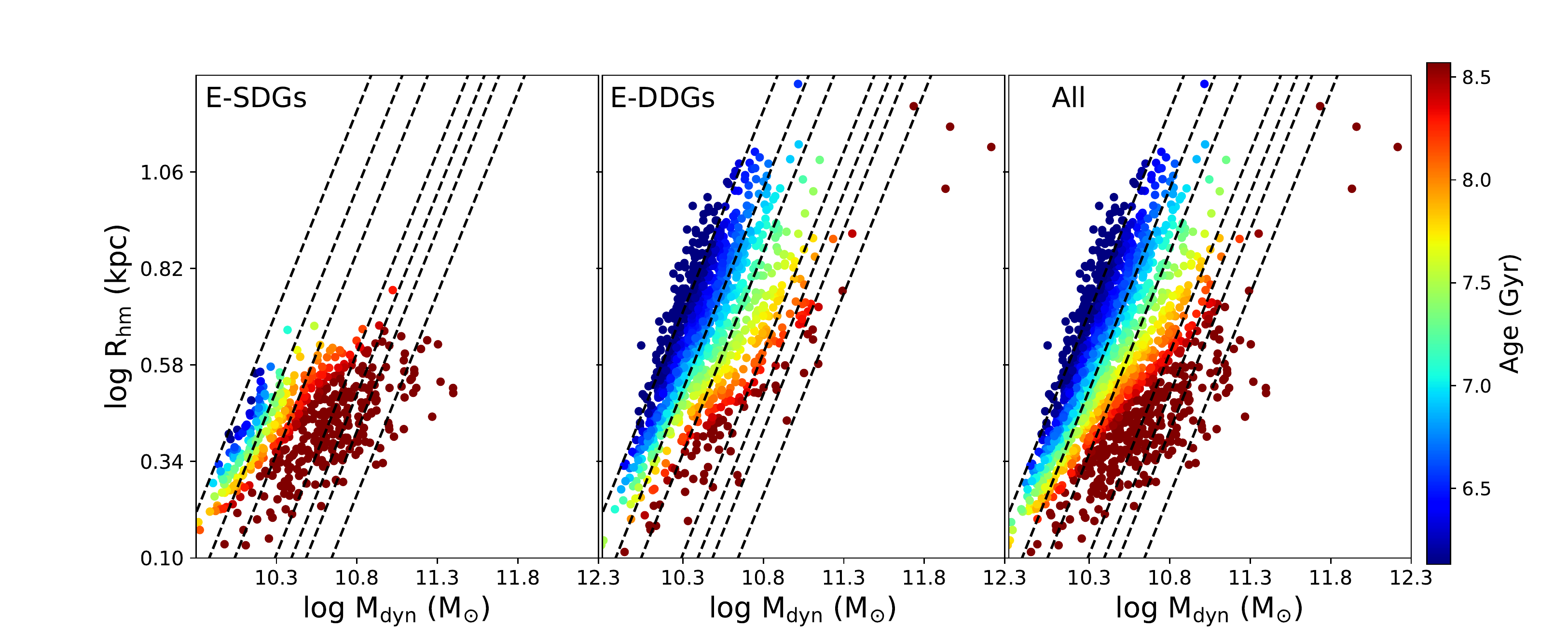}
  \caption{ Stellar mass-weighted average age  (lower panels)  LOESS-smoothed distributions on the dynamical mass-size plane for E-SDGs (left panel), E-DDGs (middle
    panel) and all galaxies (right panel) in the EAGLE simulation at $z=0$. The dashed lines show the predicted distributions for systems with constant $\sigma_{e}$
  at 100, 125, 150, 200, 225, 250 and 300 km s$^{-1}$ (from left to right).}
  \label{fig:fig1b}
\end{figure*}

Figure \ref{fig:fig1} shows the stellar mass-size plane as a function of $\sigma_{\rm e}$ for E-SDGs, E-DDGs and all galaxies together.
In this case, we consider $R_{\rm hm}^{2\mathrm{D}}$ in order to achieve a better comparison with the observations.
We use the two-dimensional Locally Weighted Regression method to obtain smoothed distributions\footnote{The colour-bars are defined by using the first and third quartile as limits.} \citep[LOESS][]{Cleveland88}. In Appendix \ref{sec:RD}, figures with the real distributions are included for comparison.
The stellar mass-size plane shows the expected trend for increasing sizes with stellar mass within the observational range reported by \citet{wel2014}. 
We find the best agreement for intermediate stellar mass galaxies $\sim 10^{10.3}$ M$_{\odot}$ for both E-SDGs and E-DDGs \citep[see also][]{sande2019}.

Globally, it can be seen that larger size galaxies have lower $\sigma_{\rm e}$ at a given stellar mass.  E-DDGs increase systematically their stellar mass and size at fixed values of $\sigma_e$. However,
$\sigma_{\rm e}$ is only a proxy of stellar mass in the case of E-SDGs.  At a similar $\sigma_{\rm e}$, E-DDGs are more extended than E-SDGs, possibly due to different assembly histories.

In Fig.~\ref{fig:fig1b}, the dynamical mass-size plane is shown  as a function of the stellar mass-weighted ages of the SPs following Li2018. The range of dynamical masses covered by the selected samples is smaller than the one provided by the observations of Li2018. 
The lack of extended massive discs as well as the narrow range of dynamical masses could be partially due to the restricted simulated volume, which is $\approx 80$ smaller than the volume covered by MaNGA.

As can be seen, galaxies with similar median stellar ages move approximately along lines of constant $\sigma_e$, in agreement with observations.
 However, the trend has a slightly flatter slope for galaxies with  $\sigma_e > 150$km s$^{-1}$. 
Young SPs dominate in galaxies with low stellar masses and $\sigma_{\rm e}$.
For a given dynamical mass, galaxies with overall old
SPs are  more compact, regardless of their morphology. These trends are in qualitative agreement with observational results  shown in  Li2018. We estimate the youngest galaxies to have median stellar ages of $\sim 3$Gyr and $\sim 3.7$ Gyr for E-DDGs and E-SDGs, and
the oldest ones of  $\sim 12.5$ for both type of galaxies. The median ages of the both samples are $\sim 6.8$ Gyr and $\sim 8.7$ Gyr,  respectively.

\begin{figure*}
  \centering
\includegraphics[width=0.9\textwidth]{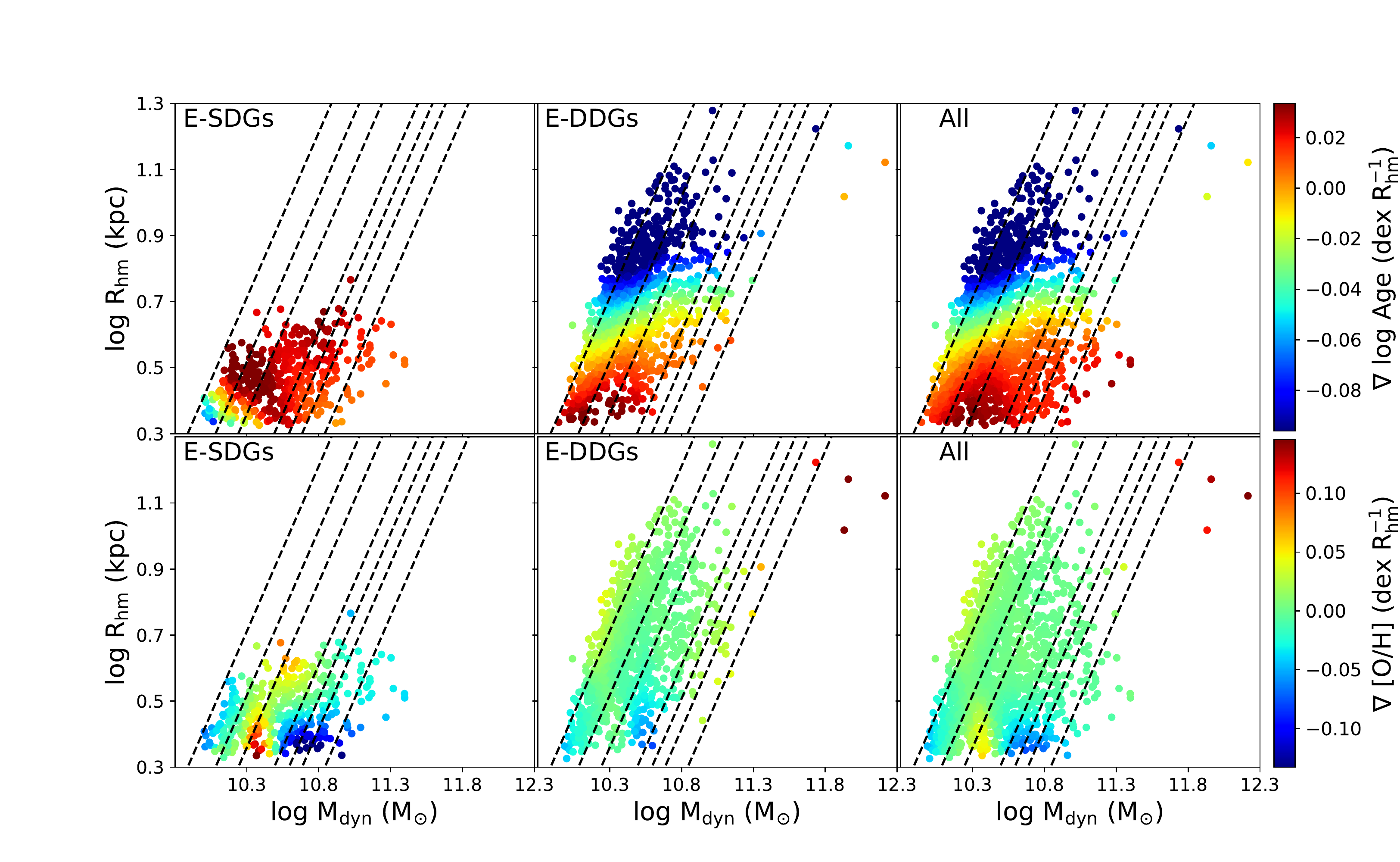}
  \caption{Age  (upper panels) and  [O/H]
     (lower panels)  gradients LOESS-smoothed  distributions on the mass-size plane for E-SDGs (left panels), E-DDGs (middle
    panels) and all galaxies (right panels)  in  EAGLE  simulation at $z=0$. Dashed lines are the same as given in Fig.~\ref{fig:fig1b}. }
  \label{fig:fig2}
\end{figure*}

Fig.~\ref{fig:fig2} shows the distribution of age and stellar metallicity gradients across the dynamical mass-size plane. 
As can be seen for the upper panels, 
massive E-SDGs  tend to have shallow/positive age gradients  which are in global qualitative agreement with Li2018. Low-mass galaxies show a larger variety of age gradients that are not reported by Li2018, probably as a result of their later assembly histories \citep[see][]{Rosito2018b}.
For E-DDGs, there is a clear trend for more extended galaxies to be more massive at a given age gradient. As the statistics are dominated by E-DDGs, when we analyse all the galaxies together, we find the trends to be similar to those of the E-DDGs. We note that the behaviour for E-DDGs is in good  agreement  with expectations of the inside-out formation model for disc galaxies.
However, we acknowledge  a lack of extended, massive galaxies in the EAGLE sample \citep{sande2019}. 

The metallicity gradients are weakly correlated with dynamical mass \citep{tissera2019}
and this is reflected on the mass-size plane  shown in the lower panels in Fig~\ref{fig:fig2}.
For the E-DDGs, metallicity gradients are globally negative.
There is a weak change in   the slopes at about $ M_{\rm dyn} \sim 10^{10.5}$M$_{\odot}$. This trend is determined mainly  by a fraction of the EAGLE
galaxies  that have  weak/positive metallicity gradients around  $\sigma_e  \sim$ 150 km s$^{-1}$. 

\begin{figure*}
  \centering
\includegraphics[width=0.9\textwidth]{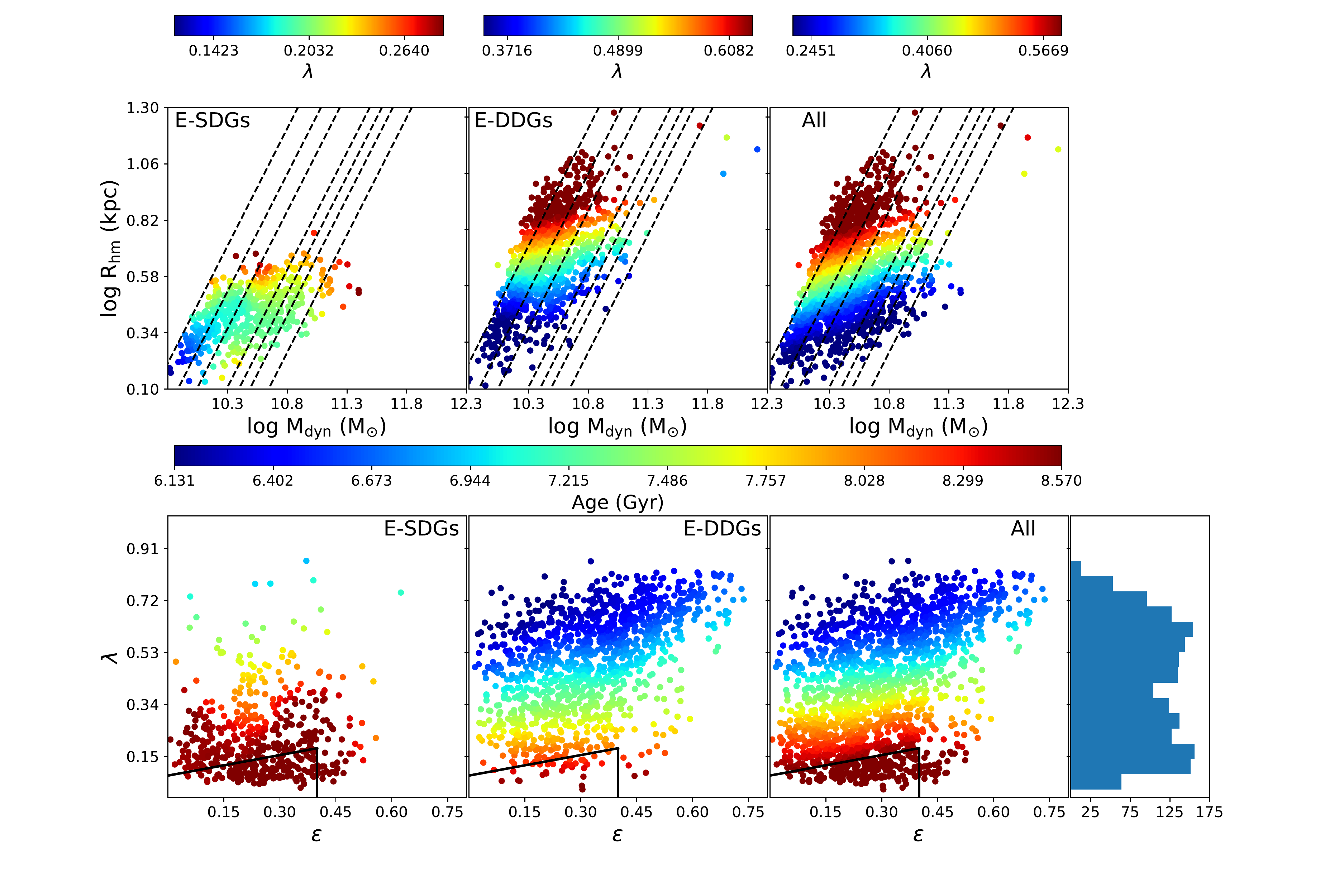}
  \caption{Upper panel: Spin parameter ($\lambda$) distribution on the mass-size plane  for E-SDGs (left panel), E-DDGs (middle
   panel) and all galaxies (right panel)  in  the  EAGLE  simulation.  Lines are the same as given in Fig.~\ref{fig:fig1b}. Lower panel: $\lambda - \epsilon$ plane coloured by stellar-mass weighted ages.
   We depict the slow rotators region according to the definition in \cite{graham2018} (black lines).
   Lower left panel: Histogram of $\lambda$ for the complete sample.}
  \label{fig:fig3}
\end{figure*}

We also analyse the mass-size plane as a function of the spin parameter $\lambda$  and the relation of the latter with $\varepsilon$ and the median stellar ages of galaxies.
In  Fig. ~\ref{fig:fig3}, we show the distribution of $\lambda$ on the dynamical mass-size plane (upper panels) in the same samples of Fig.~\ref{fig:fig1b} and  Fig.~\ref{fig:fig2} 
for comparison.
As it can be seen from this figure, there is a clear dependence on $\lambda$. Overall, E-SDGs have lower $\lambda$ than E-SDGs. For both morphological types,  at fixed $M_{\rm dyn}$,  $\lambda$ varies smoothly  with $R_{\rm hm}$ (note that different colour-bars are used in each panel in order to highlight the trends).   This figure shows that  galaxies with similar $\lambda$  are located on tracks with positive slopes in the mass-size relation, so that more massive galaxies are more extended than less massive systems at fixed $\lambda$.  
When all galaxies are analysed together, this trend remains. For the range of masses we analyse, E-SDGs tend to be more compact than E-SDGs at a given dynamical mass, with the larger differences for
the most extreme values of $\lambda$ \citep{graham2018}.

The lower panels of Fig.~\ref{fig:fig3} displays   $\lambda - \varepsilon$  as a function of the stellar mass-weighted  ages.
As can be seen the tracks of constant age are flat for $\varepsilon < \sim 0.3$, indicating  that for more oblate system, stellar age is found to be a good proxy of $\lambda$ and vice-versa. For both morphological types, older SPs are found in galaxies with
lower $\lambda$. This is consistent with these systems having experienced a larger number of (dry) mergers while those with larger $\lambda$ values have progressively younger SPs, {\rm 
consistently having with more extended star formation histories.}
We note that galaxies with $\varepsilon > \sim 0.3$  seem to be younger compared to those with smaller ellipticity at fixed $\lambda$.
However, no clean proxy for $\lambda$ is found in these galaxies.

Following previous works, if we take $\lambda \sim 0.2$ as a suitable value to separate fast and slow rotators (based on the $\lambda-\epsilon$ 
distribution of EAGLE galaxies presented; see Fig.~$3$ in \citealt{lagos2018}),  
we found that 55 per cent of the E-SDGs are slow rotators
populated by old stars on average,  while only 8 per cent of the E-DDGs are classified as slow rotators. If instead we use the definition adopted by  \citet{graham2018}, the fractions
are $\sim 28$ per cent and $\sim 2$ per cent, respectively. The fraction of fast rotator in E-SDGs is $\sim 71$ per cent, in agreement with the results reported by
\citet{lagos2018}. However this fraction is lower than that found by \cite{Emsellem2011} for ATLAS$^{3\mathrm{D}}$ ($\sim 86$ per cent) though this could be caused by ATLAS$^{3\mathrm{D}}$ selecting (by construction) ETGs only. 
Slow rotators in the EAGLE simulations are detected mostly in dispersion-dominated systems 
with old SPs. They define the second peak in the $\lambda$ histogram shown in the lower panel of  Fig.~\ref{fig:fig3}.  This histogram shows a bimodal distribution
with the first peak populated by disc galaxies (with young SPs) while the second one, by slow-rotator galaxies (with old SPs). The EAGLE galaxies are found to reproduce well the MaNGA distribution reported by \citet{graham2018}.

\section{Conclusions}
\label{sec:con}

Using a galaxy sample of simulated galaxies selected from the large $(100\,\rm Mpc)^3$ simulation of the EAGLE Project, 
we performed a statistical analysis of the dependence of the mass-size plane on a variety of galaxy properties. 
Our main results can be summarised as follow:
 
\begin{itemize}

  \item{ Early and late-type galaxies in the EAGLE simulation, so called E-SDGs and E-DDGs, are found to follow stellar mass-size planes in global agreement with observations, principally
for galaxies with stellar masses of $\sim 10^{10.3}$M$_{\odot}$. Smaller and larger mass galaxies show discrepancies but are within observed ranges reported by \citet{wel2014}.
    Larger-size galaxies have lower $\sigma_e$ values at fixed mass for simulated late-type galaxies. In the case of the simulated early-type systems, stellar mass is found to be a good proxy of
 $\sigma_e$. }

\item { Galaxies dominated by old stars are more compact at fixed dynamical mass. Li2018 report $\sigma_e$ to be a good proxy for the stellar age. We find this to be on average valid for EAGLE galaxies. However, mean stellar ages show a slightly weaker trend than $\sigma_e$ for massive disc galaxies which is not observed in Li2018.}

\item More extended discs have steeper negative age gradients while early-type galaxies show weak or more positive age gradients. The latter trend
seems to be in agreement with the slight outside-in assembly histories of low-mass,  early-type galaxies  reported by \citet{Rosito2018b}.

\item Chemical abundances are found to have a shallower dependence
on stellar mass or $\sigma_e$ than reported by observations (see Appendix \ref{app:shape}). Reproducing the observed trends with metallicity seem to be a major challenge for cosmological simulations.

\item{Overall, we find that age can be used as a proxy for $\lambda$ for more oblate systems as reported by \citet{sande2019}. Because the adopted morphological classification is based on the level of rotation  in galaxies, this implies that morphology is a good proxy for age. However, $\varepsilon$ does not show a clear trend with stellar age in the EAGLE galaxies.}

\item{ The distributions of EAGLE galaxies in the ($\lambda, \varepsilon$) plane shows  a bimodal distribution of $\lambda$ where the extended,  disc systems with young stars  populate  the first peak while the second one is dominated by slow rotator galaxies with old SPs.}

\end{itemize}

These results  suggest that the processes regulating star formation activity (and hence the age distributions) might also affect the redistribution of angular momentum of stars and gas, and hence,  galaxy morphology.  Despite the many successes of the current generation of cosmological hydrodynamical simulations, some important shortcomings remain. In the area of stellar populations, we find that some important ones are related to the metallicity profiles and age-morphology relation in galaxies. Further work on the multi-phase nature of the ISM in simulations could naturally lead to more realistic internal kinematics and profiles of galaxies and hence is key to explore.

\begin{acknowledgements}
We thank J. Trayford and J. van de Sande for their valuable comments  to this work.
PBT acknowledges partial funding by Fondecyt Regular 2015 - 1150334 and Proyecto Interno UNAB 2019.
This project has received funding from the European Union Horizon 2020
Research and Innovation Programme under the Marie Sklodowska-Curie
grant agreement No 734374 and the GALNET network funded by Conicyt.
CL is funded by the ARC Centre of
Excellence for All Sky Astrophysics in 3 Dimensions (ASTRO 3D), through project number 
CE170100013.

This work used the DiRAC Data Centric system at Durham University, operated by the Institute for Computational Cosmology on behalf of the STFC DiRAC HPC Facility (www.dirac.ac.uk). This equipment was funded by BIS National E-infrastructure capital grant ST/K00042X/1, STFC capital grants ST/H008519/1 and ST/K00087X/1, STFC DiRAC Operations grant ST/K003267/1 and Durham University. DiRAC is part of the National E-Infrastructure. We acknowledge PRACE for awarding us access to the Curie machine based in France at TGCC, CEA, Bruyeres-le-Chatel. This work used RAGNAR cluster funded by Fondecyt Regular 2015 - 1150334 at Universidad Andres Bello.

\end{acknowledgements}

\bibliographystyle{aa}
\bibliography{letter}

\begin{thebibliography}{38}
\expandafter\ifx\csname natexlab\endcsname\relax\def\natexlab#1{#1}\fi

\bibitem[{{Bois} {et~al.}(2011){Bois}, {Emsellem}, {Bournaud}, {Alatalo},
  {Blitz}, {Bureau}, {Cappellari}, {Davies}, {Davis}, {de Zeeuw}, {Duc},
  {Khochfar}, {Krajnovi{\'c}}, {Kuntschner}, {Lablanche}, {McDermid},
  {Morganti}, {Naab}, {Oosterloo}, {Sarzi}, {Scott}, {Serra}, {Weijmans}, \&
  {Young}}]{bois2011}
{Bois}, M., {Emsellem}, E., {Bournaud}, F., {et~al.} 2011, \mnras, 416, 1654

\bibitem[{{Bryant} {et~al.}(2015){Bryant}, {Owers}, {Robotham}, {Croom},
  {Driver}, \& {Drinkwater}}]{bryant2015}
{Bryant}, J.~J., {Owers}, M.~S., {Robotham}, A.~S.~G., {et~al.} 2015, \mnras,
  447, 2857

\bibitem[{{Bundy} {et~al.}(2015){Bundy}, {Bershady}, {Law}, {Yan}, {Drory},
  {MacDonald}, {Wake}, {Cherinka}, {S{\'a}nchez-Gallego}, {Weijmans}, \&
  {Thomas}}]{Bundy2015}
{Bundy}, K., {Bershady}, M.~A., {Law}, D.~R., {et~al.} 2015, \apj, 798, 7

\bibitem[{{Cappellari} {et~al.}(2013){Cappellari}, {McDermid}, {Alatalo},
  {Blitz}, {Bois}, {Bournaud}, {Bureau}, {Crocker}, {Davies}, {Davis}, \& {et
  al.}}]{CappellariAtlasXX}
{Cappellari}, M., {McDermid}, R.~M., {Alatalo}, K., {et~al.} 2013, \mnras, 432,
  1862

\bibitem[{{Clauwens} {et~al.}(2018){Clauwens}, {Schaye}, {Franx}, \&
  {Bower}}]{clauwens2018}
{Clauwens}, B., {Schaye}, J., {Franx}, M., \& {Bower}, R.~G. 2018, \mnras, 478,
  3994

\bibitem[{Cleveland \& Devlin(1988)}]{Cleveland88}
Cleveland, W.~S. \& Devlin, S.~J. 1988, Journal of the American Statistical
  Association, 83, 596

\bibitem[{{Crain} {et~al.}(2015){Crain}, {Schaye}, {Bower}, {Furlong},
  {Schaller}, {Theuns}, {Dalla Vecchia}, {Frenk}, {McCarthy}, {Helly},
  {Jenkins}, {Rosas-Guevara}, {White}, \& {Trayford}}]{Crain2015}
{Crain}, R.~A., {Schaye}, J., {Bower}, R.~G., {et~al.} 2015, \mnras, 450, 1937

\bibitem[{{De Rossi} {et~al.}(2017){De Rossi}, {Bower}, {Font}, {Schaye}, \&
  {Theuns}}]{derossi2017}
{De Rossi}, M.~E., {Bower}, R.~G., {Font}, A.~S., {Schaye}, J., \& {Theuns}, T.
  2017, \mnras, 472, 3354

\bibitem[{{Emsellem} {et~al.}(2011){Emsellem}, {Cappellari}, {Krajnovi{\'c}},
  {Alatalo}, {Blitz}, {Bois}, {Bournaud}, {Bureau}, {Davies}, {Davis}, {de
  Zeeuw}, {Khochfar}, {Kuntschner}, {Lablanche}, {McDermid}, {Morganti},
  {Naab}, {Oosterloo}, {Sarzi}, {Scott}, {Serra}, {van de Ven}, {Weijmans}, \&
  {Young}}]{Emsellem2011}
{Emsellem}, E., {Cappellari}, M., {Krajnovi{\'c}}, D., {et~al.} 2011, \mnras,
  414, 888

\bibitem[{{Fall} \& {Romanowsky}(2013)}]{fall2013}
{Fall}, S.~M. \& {Romanowsky}, A.~J. 2013, \apjl, 769, L26

\bibitem[{Frenk {et~al.}(2018)Frenk, Theuns, Trayford, Correa, \&
  Schaye}]{Trayford2018}
Frenk, C.~S., Theuns, T., Trayford, J.~W., Correa, C., \& Schaye, J. 2018,
  \mnras, 483, 744

\bibitem[{{Graham} {et~al.}(2018){Graham}, {Cappellari}, {Li}, {Mao},
  {Bershady}, {Bizyaev}, {Brinkmann}, {Brownstein}, {Bundy}, {Drory}, {Law},
  {Pan}, {Thomas}, {Wake}, {Weijmans}, {Westfall}, \& {Yan}}]{graham2018}
{Graham}, M.~T., {Cappellari}, M., {Li}, H., {et~al.} 2018, \mnras, 477, 4711

\bibitem[{{Lagos} {et~al.}(2018{\natexlab{a}}){Lagos}, {Schaye}, {Bah{\'e}},
  {Van de Sande}, {Kay}, {Barnes}, {Davis}, \& {Dalla Vecchia}}]{lagos2018}
{Lagos}, C.~d.~P., {Schaye}, J., {Bah{\'e}}, Y., {et~al.} 2018{\natexlab{a}},
  \mnras, 476, 4327

\bibitem[{{Lagos} {et~al.}(2018{\natexlab{b}}){Lagos}, {Stevens}, {Bower},
  {Davis}, {Contreras}, {Padilla}, {Obreschkow}, {Croton}, {Trayford},
  {Welker}, \& {Theuns}}]{Lagos2018b}
{Lagos}, C.~d.~P., {Stevens}, A.~R.~H., {Bower}, R.~G., {et~al.}
  2018{\natexlab{b}}, \mnras, 473, 4956

\bibitem[{{Lagos} {et~al.}(2017){Lagos}, {Theuns}, {Stevens}, {Cortese},
  {Padilla}, {Davis}, {Contreras}, \& {Croton}}]{lagos2017}
{Lagos}, C.~d.~P., {Theuns}, T., {Stevens}, A.~R.~H., {et~al.} 2017, \mnras,
  464, 3850

\bibitem[{{Li} {et~al.}(2018){Li}, {Mao}, {Cappellari}, {Ge}, {Long}, {Li},
  {Mo}, {Li}, {Zheng}, {Bundy}, {Thomas}, {Brownstein}, {Roman Lopes}, {Law},
  \& {Drory}}]{Li2018}
{Li}, H., {Mao}, S., {Cappellari}, M., {et~al.} 2018, \mnras, 476, 1765

\bibitem[{{Lia} {et~al.}(2002){Lia}, {Portinari}, \& {Carraro}}]{lia2002}
{Lia}, C., {Portinari}, L., \& {Carraro}, G. 2002, \mnras, 330, 821

\bibitem[{{McAlpine} {et~al.}(2016){McAlpine}, {Helly}, {Schaller}, {Trayford},
  {Qu}, {Furlong}, {Bower}, {Crain}, {Schaye}, {Theuns}, {Dalla Vecchia},
  {Frenk}, {McCarthy}, {Jenkins}, {Rosas-Guevara}, {White}, {Baes}, {Camps}, \&
  {Lemson}}]{mcalpine2016}
{McAlpine}, S., {Helly}, J.~C., {Schaller}, M., {et~al.} 2016, Astronomy and
  Computing, 15, 72

\bibitem[{{Mosconi} {et~al.}(2001){Mosconi}, {Tissera}, {Lambas}, \&
  {Cora}}]{mosco2001}
{Mosconi}, M.~B., {Tissera}, P.~B., {Lambas}, D.~G., \& {Cora}, S.~A. 2001,
  \mnras, 325, 34

\bibitem[{{Naab}(2013)}]{Naab2013}
{Naab}, T. 2013, in IAU Symposium, Vol. 295, The Intriguing Life of Massive
  Galaxies, ed. D.~{Thomas}, A.~{Pasquali}, \& I.~{Ferreras}, 340--349

\bibitem[{{Naab} {et~al.}(2014){Naab}, {Oser}, {Emsellem}, {Cappellari},
  {Krajnovi{\'c}}, {McDermid}, {Alatalo}, {Bayet}, {Blitz}, {Bois}, {Bournaud},
  {Bureau}, {Crocker}, {Davies}, {Davis}, {de Zeeuw}, {Duc}, {Hirschmann},
  {Johansson}, {Khochfar}, {Kuntschner}, {Morganti}, {Oosterloo}, {Sarzi},
  {Scott}, {Serra}, {van de Ven}, {Weijmans}, \& {Young}}]{naab2014}
{Naab}, T., {Oser}, L., {Emsellem}, E., {et~al.} 2014, \mnras, 444, 3357

\bibitem[{{Pedrosa} \& {Tissera}(2015)}]{Pedrosa2015}
{Pedrosa}, S.~E. \& {Tissera}, P.~B. 2015, \aap, 584, A43

\bibitem[{{Perez} {et~al.}(2013){Perez}, {Valenzuela}, {Tissera}, \&
  {Michel-Dansac}}]{perez2013}
{Perez}, J., {Valenzuela}, O., {Tissera}, P.~B., \& {Michel-Dansac}, L. 2013,
  \mnras, 436, 259

\bibitem[{{Planck Collaboration} {et~al.}(2014{\natexlab{a}}){Planck
  Collaboration}, {Ade}, {Aghanim}, {Alves}, {Armitage-Caplan}, {Arnaud},
  {Ashdown}, {Atrio-Barandela}, {Aumont}, {Aussel}, \& et~al.}]{Planck2014a}
{Planck Collaboration}, {Ade}, P.~A.~R., {Aghanim}, N., {et~al.}
  2014{\natexlab{a}}, \aap, 571, A1

\bibitem[{{Planck Collaboration} {et~al.}(2014{\natexlab{b}}){Planck
  Collaboration}, {Ade}, {Aghanim}, {Armitage-Caplan}, {Arnaud}, {Ashdown},
  {Atrio-Barandela}, {Aumont}, {Baccigalupi}, {Banday}, \&
  et~al.}]{Planck2014b}
{Planck Collaboration}, {Ade}, P.~A.~R., {Aghanim}, N., {et~al.}
  2014{\natexlab{b}}, \aap, 571, A16

\bibitem[{{Rosas-Guevara} {et~al.}(2016){Rosas-Guevara}, {Bower}, {Schaye},
  {McAlpine}, {Dalla Vecchia}, {Frenk}, {Schaller}, \& {Theuns}}]{Rosas2016}
{Rosas-Guevara}, Y., {Bower}, R.~G., {Schaye}, J., {et~al.} 2016, \mnras, 462,
  190

\bibitem[{{Rosito} {et~al.}(2018){Rosito}, {Tissera}, {Pedrosa}, \&
  {Rosas-Guevara}}]{Rosito2018b}
{Rosito}, M.~S., {Tissera}, P.~B., {Pedrosa}, S.~E., \& {Rosas-Guevara}, Y.
  2018, ArXiv e-prints [\eprint[arXiv]{1811.11062}]

\bibitem[{{Ryden} {et~al.}(2001){Ryden}, {Forbes}, \& {Terlevich}}]{Ryden2001}
{Ryden}, B.~S., {Forbes}, D.~A., \& {Terlevich}, A.~I. 2001, \mnras, 326, 1141

\bibitem[{{S{\'a}nchez} {et~al.}(2012){S{\'a}nchez}, {Kennicutt}, {Gil de Paz},
  {van de Ven}, {V{\'{\i}}lchez}, {Wisotzki}, {Walcher}, {Mast}, {Aguerri},
  {Albiol-P{\'e}rez}, \& {Alonso-Herrero}}]{CALIFA2012}
{S{\'a}nchez}, S.~F., {Kennicutt}, R.~C., {Gil de Paz}, A., {et~al.} 2012,
  \aap, 538, A8

\bibitem[{{Schaye} {et~al.}(2015){Schaye}, {Crain}, {Bower}, {Furlong},
  {Schaller}, {Theuns}, {Dalla Vecchia}, {Frenk}, {McCarthy}, {Helly},
  {Jenkins}, {Rosas-Guevara}, {White}, {Baes}, {Booth}, {Camps}, {Navarro},
  {Qu}, {Rahmati}, {Sawala}, {Thomas}, \& {Trayford}}]{schaye2015}
{Schaye}, J., {Crain}, R.~A., {Bower}, R.~G., {et~al.} 2015, \mnras, 446, 521

\bibitem[{{Scott} {et~al.}(2017){Scott}, {Brough}, {Croom}, {Davies}, {van de
  Sande}, {Allen}, {Bland-Hawthorn}, {Bryant}, {Cortese}, {D'Eugenio},
  {Federrath}, {Ferreras}, {Goodwin}, {Groves}, {Konstantopoulos}, {Lawrence},
  {Medling}, {Moffett}, {Owers}, {Richards}, {Robotham}, {Tonini}, \&
  {Yi}}]{Scott2017}
{Scott}, N., {Brough}, S., {Croom}, S.~M., {et~al.} 2017, \mnras, 472, 2833

\bibitem[{{Taylor} \& {Kobayashi}(2017)}]{taylor2017}
{Taylor}, P. \& {Kobayashi}, C. 2017, \mnras, 471, 3856

\bibitem[{{Tissera} {et~al.}(2016){Tissera}, {Machado}, {Sanchez-Blazquez},
  {Pedrosa}, {S{\'a}nchez}, {Snaith}, \& {Vilchez}}]{Tissera2016}
{Tissera}, P.~B., {Machado}, R.~E.~G., {Sanchez-Blazquez}, P., {et~al.} 2016,
  \aap, 592, A93

\bibitem[{{Tissera} {et~al.}(2019){Tissera}, {Rosas-Guevara}, {Bower}, {Crain},
  {del P Lagos}, {Schaller}, {Schaye}, \& {Theuns}}]{tissera2019}
{Tissera}, P.~B., {Rosas-Guevara}, Y., {Bower}, R.~G., {et~al.} 2019, \mnras,
  482, 2208

\bibitem[{{Tissera} {et~al.}(2012){Tissera}, {White}, \&
  {Scannapieco}}]{tissera2012}
{Tissera}, P.~B., {White}, S.~D.~M., \& {Scannapieco}, C. 2012, \mnras, 420,
  255

\bibitem[{{van de Sande} {et~al.}(2019){van de Sande}, {Lagos}, {Welker},
  {Bland-Hawthorn}, {Schulze}, {Remus}, {Bah{\'e}}, {Brough}, {Bryant},
  {Cortese}, {Croom}, {Devriendt}, {Dubois}, {Goodwin}, {Konstantopoulos},
  {Lawrence}, {Medling}, {Pichon}, {Richards}, {Sanchez}, {Scott}, \&
  {Sweet}}]{sande2019}
{van de Sande}, J., {Lagos}, C.~D.~P., {Welker}, C., {et~al.} 2019, \mnras,
  484, 869

\bibitem[{{van de Sande} {et~al.}(2018){van de Sande}, {Scott},
  {Bland-Hawthorn}, {Brough}, {Bryant}, {Colless}, {Cortese}, {Croom},
  {d'Eugenio}, {Foster}, {Goodwin}, {Konstantopoulos}, {Lawrence}, {McDermid},
  {Medling}, {Owers}, {Richards}, \& {Sharp}}]{sande2018}
{van de Sande}, J., {Scott}, N., {Bland-Hawthorn}, J., {et~al.} 2018, Nature
  Astronomy, 2, 483

\bibitem[{{van der Wel} {et~al.}(2014){van der Wel}, {Franx}, {van Dokkum},
  {Skelton}, {Momcheva}, {Whitaker}, {Brammer}, {Bell}, {Rix}, {Wuyts},
  {Ferguson}, {Holden}, {Barro}, {Koekemoer}, {Chang}, {McGrath},
  {H{\"a}ussler}, {Dekel}, {Behroozi}, {Fumagalli}, {Leja}, {Lundgren},
  {Maseda}, {Nelson}, {Wake}, {Patel}, {Labb{\'e}}, {Faber}, {Grogin}, \&
  {Kocevski}}]{wel2014}
{van der Wel}, A., {Franx}, M., {van Dokkum}, P.~G., {et~al.} 2014, \apj, 788,
  28

\end{thebibliography}

\begin{appendix} 

\section{ Supplementary information on EAGLE galaxies}
\label{app:shape}

\subsection{The mass-size plane as a function of metallicity}

\begin{figure*}
  \centering
\includegraphics[width=0.9\textwidth]{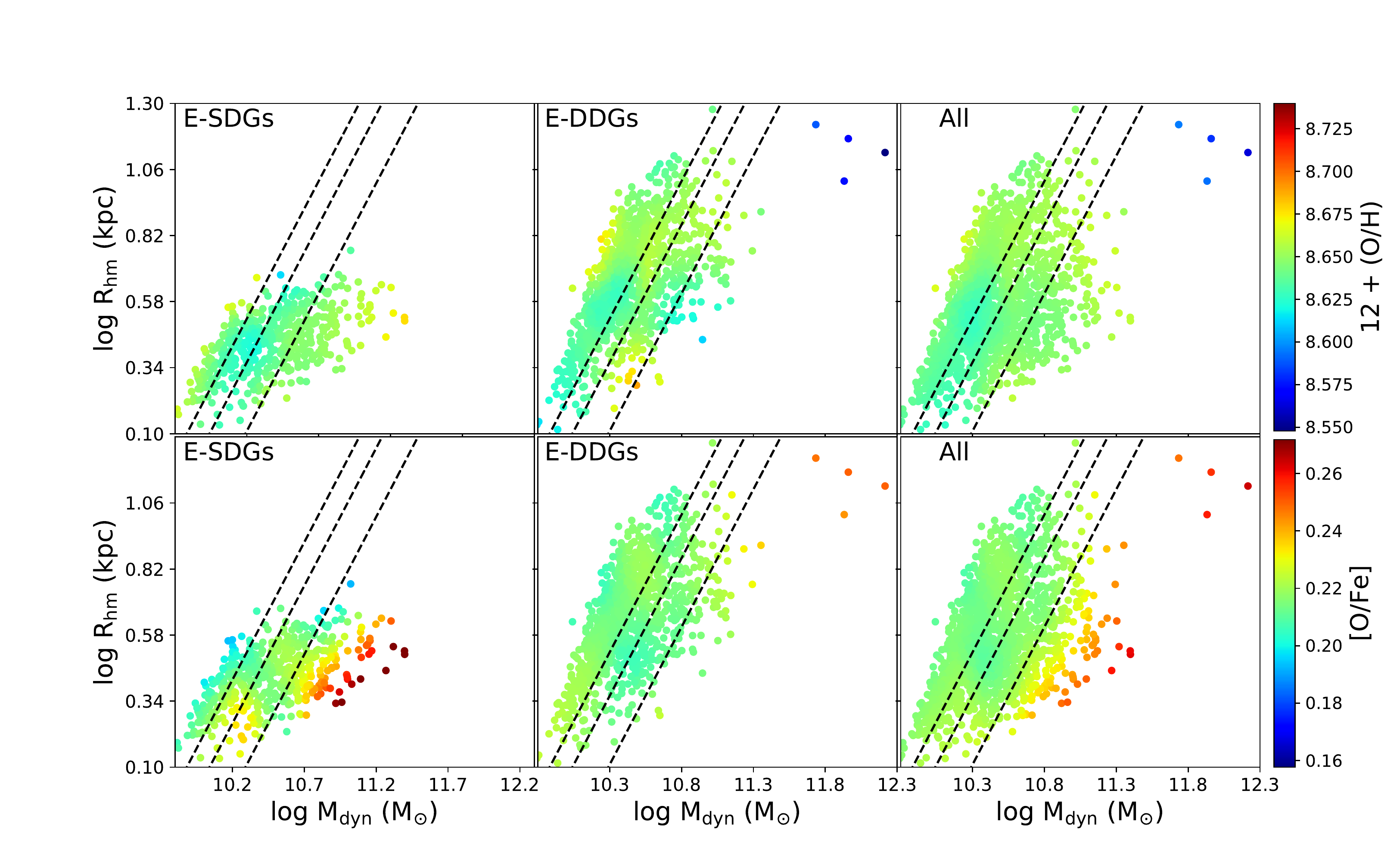}
  \caption{Median (O/H)+12 (upper panels) and 
    [O/Fe] (lower panels) LOESS-smoothed  distributions on the mass-size plane for E-SDGs (left panel), E-DDGs (middle
    panel)   and all galaxies (right panel) in the EAGLE sample. Lines are the same as given in Fig.~\ref{fig:fig1b}. }
  \label{fig:a1}
\end{figure*}

As a function of chemical abundances, we find very weak trends either using (O/H) or [O/Fe]  as shown in Fig. \ref{fig:a1}. This is consistent with previous results  \citep{derossi2017, tissera2019}.
There are also quite  large dispersions in the values which make difficult disentangle clear trends (see also Tissera \& Solar in preparation for a discussion on the metallicity dispersion). 
The [O/Fe] distribution shows  a slight trend  for massive E-SDGs to have higher values. This is expected since  massive E-SDGs would tend to form stars in  short and strong starbursts \citep{Rosito2018b}, and hence   the chemical abundances are mainly determined by SNe II nucleosynthesis \citep{Scott2017}. 
It is also interesting to note that, at a given stellar mass, there is a trend for more massive galaxies to have lower [O/Fe]. Nevertheless, we point out for both type of galaxies  no systematic
dependencies of chemical abundances with mass or $\sigma_e$ are detected.

\section{ Real distributions}
\label{sec:RD}
In this section, we include the collection of figures shown in the main paper and in Appendix \ref{app:shape}, using the real distributions instead of LOESS-smoothed ones. 
This set of figures are given only for comparison.

\begin{figure*}
  \centering
\includegraphics[width=0.9\textwidth]{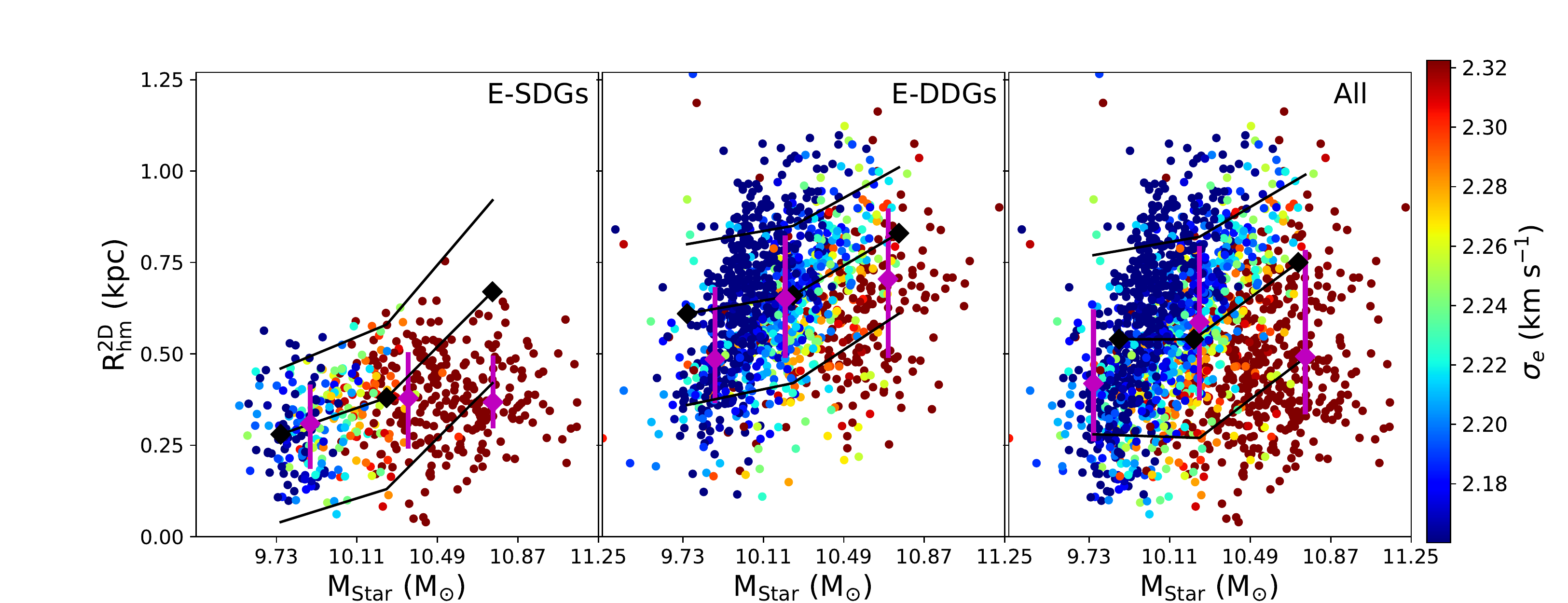}
  \caption{Velocity dispersion ($\sigma_e$) LOESS-smoothed distributions  on the stellar mass-size plane for E-SDGs (left panel), E-DDGs (middle
    panel) and all galaxies (right panel) in the EAGLE simulation at $z=0$. The median relations for the EAGLE galaxies (pink rhombus) are also shown.
For comparison the median relations for passive (left panel), active (middle panel) and all galaxies together (right panel) reported by \citet{wel2014} are also included in black rhombus 
 Error bars correspond to the 16 and 84 percentiles  for both 
simulated and observed data. See Fig. \ref{fig:fig1} for the smoothed version.}
  \label{fig:fig1_ns}
\end{figure*}

\begin{figure*}
  \centering
\includegraphics[width=0.9\textwidth]{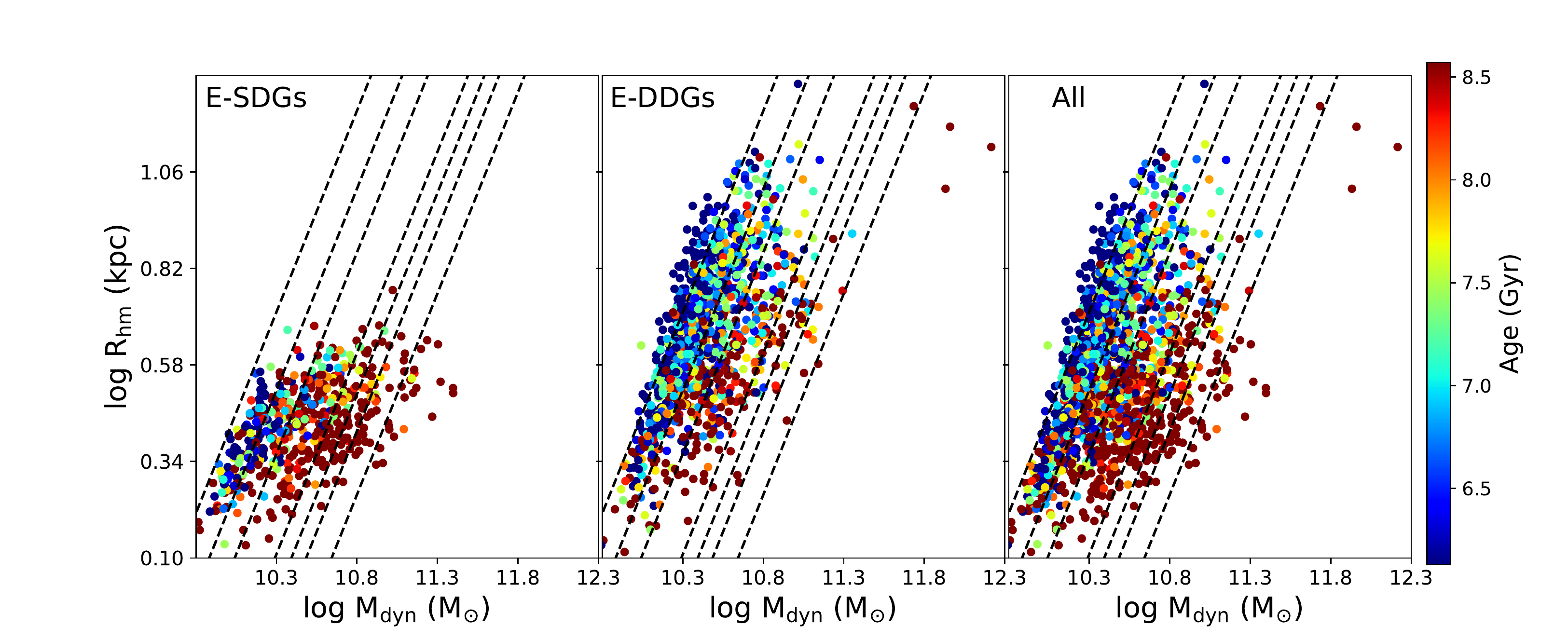}
  \caption{ Stellar mass-weighted average age  (lower panels)  LOESS-smoothed distributions on the dynamical mass-size plane for E-SDGs (left panel), E-DDGs (middle
    panel) and all galaxies (right panel) in the EAGLE simulation at $z=0$. The dashed lines show the predicted distributions for systems with constant $\sigma_{e}$
  at 100, 125, 150, 200, 225, 250 and 300 km s$^{-1}$ (from left to right).
  See Fig. \ref{fig:fig1b} for the smoothed version.}
  \label{fig:fig1b_ns}
\end{figure*}

\begin{figure*}
  \centering
\includegraphics[width=0.9\textwidth]{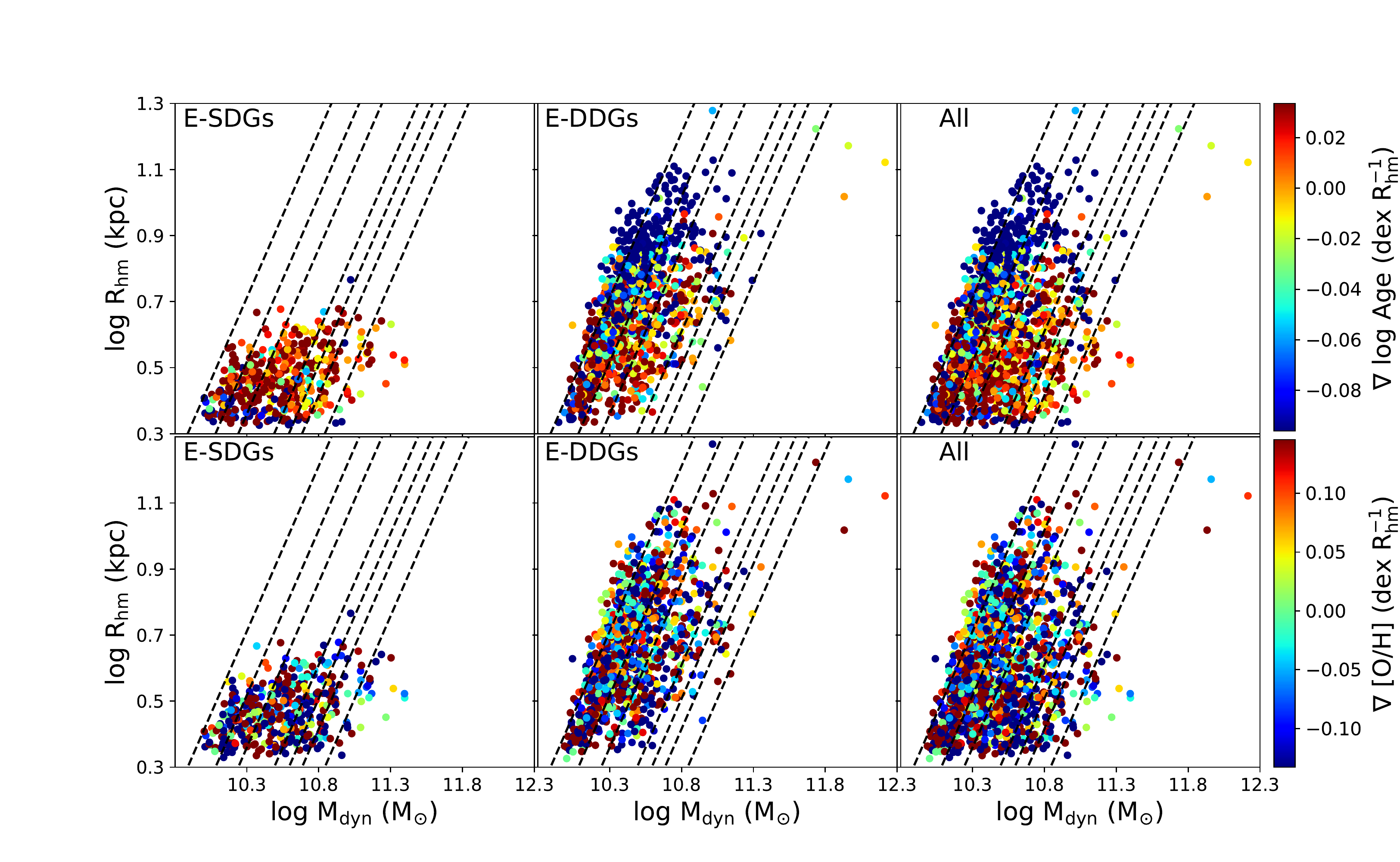}
  \caption{Age  (upper panels) and  [O/H]
     (lower panels)  gradients LOESS-smoothed  distributions on the mass-size plane for E-SDGs (left panels), E-DDGs (middle
    panels) and all galaxies (right panels)  in  EAGLE  simulation at $z=0$. Dashed lines are the same as given in Fig.~\ref{fig:fig1b}. 
    See Fig. \ref{fig:fig2} for the smoothed version.}
  \label{fig:fig2_ns}
\end{figure*}

\begin{figure*}
  \centering
\includegraphics[width=0.9\textwidth]{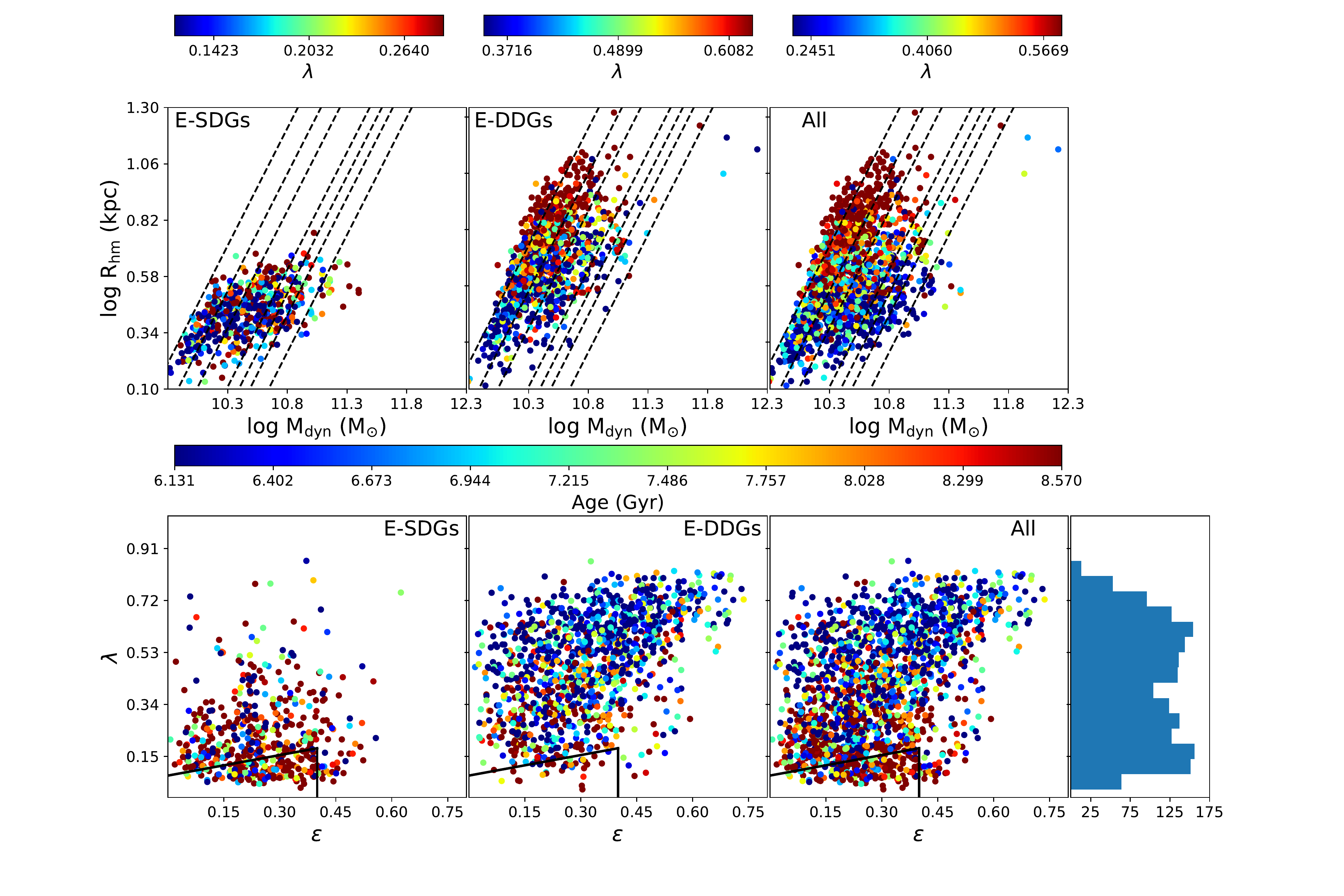}
  \caption{Upper panel: Spin parameter ($\lambda$) distribution on the mass-size plane  for E-SDGs (left panel), E-DDGs (middle
   panel) and all galaxies (right panel)  in  the  EAGLE  simulation.  Lines are the same as given in Fig.~\ref{fig:fig1b}. Lower panel: $\lambda - \epsilon$ plane coloured by stellar-mass weighted ages.
      We depict the slow rotators region according to the definition in \cite{graham2018} (black lines).
   Lower left panel: Histogram of $\lambda$ for the complete sample.
   See Fig. \ref{fig:fig3} for the smoothed version.}
  \label{fig:fig3_ns}
\end{figure*}

\begin{figure*}
  \centering
\includegraphics[width=0.9\textwidth]{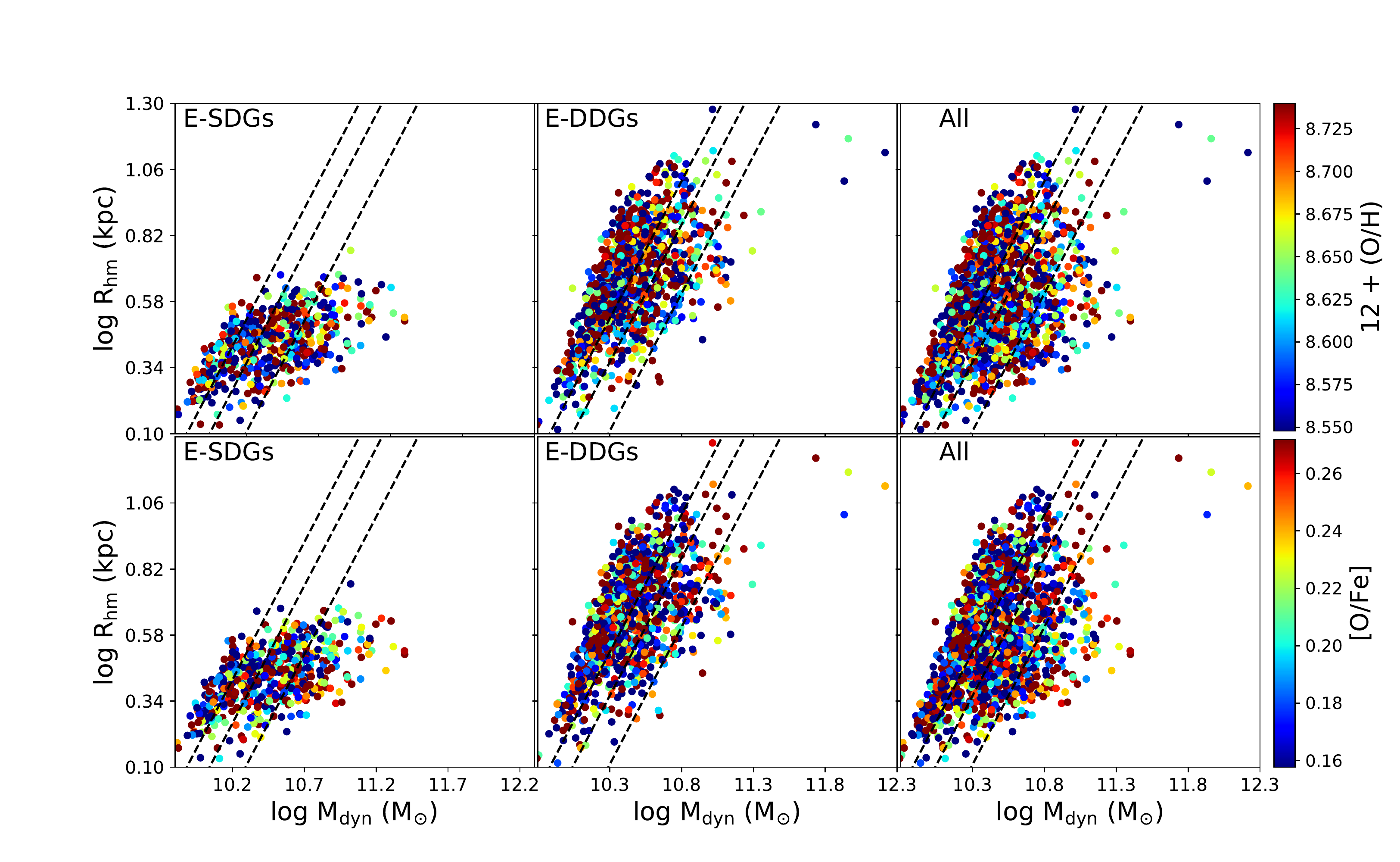}
  \caption{Median (O/H)+12 (upper panels) and 
    [O/Fe] (lower panels) distributions on the mass-size plane for E-SDGs (left panel), E-DDGs (middle
    panel)   and all galaxies (right panel) in the EAGLE sample. Lines are as given in Fig.~\ref{fig:fig1}. 
    See Fig. \ref{fig:a1} for the smoothed version.}
  \label{fig:A1NS}
\end{figure*}

\end{appendix}

\end{document}